**A "toy" model in QFT with no lower bound to the energy**

Dan Solomon
Rauland-Borg Corporation
Mount Prospect, IL

Email: **dan.solomon@rauland.com**

(March 14, 2009)

**Abstract**

In quantum field theory, it is generally assumed that there is a lower bound to the energy which is normally assumed to be the vacuum state. While this may be a reasonable assumption for a free field it is not necessarily the case for interacting fields. In this paper I will examine a "toy" model of a neutral scalar field interacting with a charged scalar field and show that there is no lower bound to the energy for this case.

## 1. Introduction.

It is generally assumed, in quantum field theory, that there is a lower bound to the energy. While this may be a reasonable assumption for a free field it has been shown that this is not necessarily the case for interacting fields. For example it has been shown that for a Dirac-Maxwell field in the temporal gauge there is no lower bound to the energy [1][2]. This has also been shown to be true for a Dirac-Maxwell field in the covariant gauge[3]. Here we examine a "toy" model of an interacting field and show in this case there is no lower bound to the energy. The model shall be formulated in 1-1D space-time where $x$ is the space dimension and $t$ is time.

Consider a system consisting of a neutral scalar field coupled to a charged scalar field. Let the Hamiltonian be given by,

$$\hat{H} = \hat{H}_{0,s} + \hat{H}_{0,cs} + \lambda_1 \int :\hat{\phi}^\dagger \hat{\phi}: \hat{\varphi} dx + \lambda_2 \int :\hat{\varphi}^4: dx \qquad (1.1)$$

where $\hat{\varphi}(x)$ is the field operator for the neutral scalar field, $\hat{\phi}(x)$ is the field operator of the charged scalar field, $\hat{H}_{0,s}$ is the Hamiltonian for the free neutral scalar field, $\hat{H}_{0,cs}$ is the Hamiltonian for the free charged scalar field, the $\lambda_1$ and $\lambda_2$ are positive constants, and an operator with colons, e.g., $:\hat{\varphi}^4:$, denotes normal order.

The field operator $\hat{\varphi}(x)$ is given by,

$$\hat{\varphi}(x) = \sum_p \frac{1}{\sqrt{2\omega_p L}} \left( \hat{a}_p e^{ipx} + \hat{a}_p^\dagger e^{-ipx} \right) \qquad (1.2)$$

In the above equation $L$ is the one-dimensional integration volume, $\omega_p = \sqrt{p^2 + m^2}$ where $m$ is the mass of the neutral scalar field, $p = 2\pi n/L$ where $n$ is an integer, and the $\hat{a}_p$ and $\hat{a}_p^\dagger$ satisfy the usual commutation relationship,

$$\left[ \hat{a}_p, \hat{a}_q^\dagger \right] = \delta_{pq} \qquad (1.3)$$

The field operator for the charged scalar field is,

$$\hat{\phi}(x) = \sum_p \frac{1}{\sqrt{2E_p L}} \left( \hat{b}_p e^{ipx} + \hat{d}_p^\dagger e^{-ipx} \right) \qquad (1.4)$$

where the complex conjugate operator is,

$$\phi^\dagger(x) = \sum_p \frac{1}{\sqrt{2E_p L}} \left( \hat{b}_p^\dagger e^{-ipx} + \hat{d}_p e^{ipx} \right) \tag{1.5}$$

where $E_p = \sqrt{p^2 + M^2}$ with $M$ being the mass of the charged scalar field. The usual commutation relationships hold,

$$\left[ \hat{b}_p, \hat{b}_q^\dagger \right] = \delta_{pq}; \quad \left[ \hat{d}_p, \hat{d}_q^\dagger \right] = \delta_{pq} \tag{1.6}$$

with all other commutations being equal to zero. Note that all commutations of any of the $\hat{a}_p$ and $\hat{a}_p^\dagger$ with the $\hat{b}_q$, $\hat{b}_q^\dagger$, $\hat{d}_q$, or $\hat{d}_q^\dagger$ are zero.

The free field part of the Hamiltonian is given by,

$$\hat{H}_{0,cs} = \sum_p E_p \left( b_p^\dagger b_p + d_p^\dagger d_p \right) \tag{1.7}$$

and,

$$\hat{H}_{0,s} = \sum_p \omega_p a_p^\dagger a_p \tag{1.8}$$

## 2. The energy of a state.

The energy of a normalized state vector $|\Omega\rangle$ is defined as $\langle \Omega | \hat{H} | \Omega \rangle$. The question we want to address is whether or not there is a lower bound to this energy. First, assume that there exists a normalized state $|\Omega_v\rangle$ which is the lowest energy state, i.e., if $|\Omega\rangle$ is a normalized state vector then,

$$\langle \Omega | \hat{H} | \Omega \rangle \geq \langle \Omega_v | \hat{H} | \Omega_v \rangle \text{ for all } |\Omega\rangle \tag{1.9}$$

In order to test this assumption we will try to find a state vector that violates this relationship. Define the operators

$$\hat{U}_{cs} = \exp\left\{ f_1 \left( \left( \hat{b}_q^\dagger + \hat{d}_q^\dagger \right) - \left( \hat{b}_q + \hat{d}_q \right) \right) \right\} \text{ and } \hat{U}_s = \exp\left( f_2 \left( \hat{a}_k^\dagger - \hat{a}_k \right) \right) \tag{1.10}$$

where $f_1$ and $f_2$ are real valued constants. It is evident that $\hat{U}_{cs}^\dagger = \hat{U}_{cs}^{-1}$ and $\hat{U}_s^\dagger = \hat{U}_s^{-1}$ therefore $\hat{U}_{cs}$ and $\hat{U}_s$ are unitary operators. In addition $\hat{U}_{cs}$ commutes with $\hat{U}_s$. Next define $\hat{U} = \hat{U}_{cs} \hat{U}_s$. From the above it is evident that $\hat{U}^\dagger = \hat{U}^{-1}$ so that $\hat{U}$ is a unitary operator.

The Baker-Campell-Hausdorff relationship is given by,

$$e^{-A}Be^{A} = B + [B,A] + \frac{1}{2}[[B,A],A] + \ldots \qquad (1.11)$$

Using this relationship and the commutation relationships, (1.3) and (1.6), we obtain,

$$\hat{U}^\dagger \hat{b}_p \hat{U} = \hat{b}_p + \delta_{pq} f_1; \quad \hat{U}^\dagger \hat{b}_p^\dagger \hat{U} = \hat{b}_p^\dagger + \delta_{pq} f_1; \quad \hat{U}^\dagger \hat{d}_p \hat{U} = \hat{d}_p + \delta_{qp} f_1; \quad \hat{U}^\dagger \hat{d}_p^\dagger \hat{U} = \hat{d}_p^\dagger + \delta_{pq} f_1$$
$$(1.12)$$

and,

$$\hat{U}^\dagger \hat{a}_p \hat{U} = \hat{a}_p + \delta_{pk} f_2; \quad \hat{U}^\dagger \hat{a}_p^\dagger \hat{U} = \hat{a}_p^\dagger + \delta_{pk} f_2 \qquad (1.13)$$

Use these relationships, and the fact that, $\hat{U}^\dagger \hat{U} = 1$ to obtain,

$$\hat{U}^\dagger \hat{H}_{0,s} U = \sum_p \omega_p \hat{U}^\dagger \hat{a}_p^\dagger \hat{U} \hat{U}^\dagger \hat{a}_p \hat{U}^\dagger = \sum_{p \ne k} \omega_p \hat{a}_p^\dagger \hat{a}_p + \omega_k \left( \hat{a}_k^\dagger + f_2 \right)\left( \hat{a}_k + f_2 \right) \qquad (1.14)$$

Rearrange terms to obtain,

$$\hat{U}^\dagger \hat{H}_{0,s} U = \hat{H}_{0,s} + \omega_k \left[ f_2 \left( \hat{a}_k^\dagger + \hat{a}_k \right) + f_2^2 \right] \qquad (1.15)$$

Similarly, we can show that,

$$\hat{U}^\dagger \hat{H}_{0,cs} U = \hat{H}_{0,cs} + E_q \left[ f_1 \left( \hat{b}_q^\dagger + \hat{b}_q + \hat{d}_q^\dagger + \hat{d}_q \right) + 2 f_1^2 \right] \qquad (1.16)$$

Next use (1.12) and (1.13) to obtain the following relationships,

$$\hat{U}^\dagger \hat{\varphi}(x) \hat{U} = \hat{\varphi}(x) + \frac{f_2}{\sqrt{2\omega_k L}} \left( e^{ikx} + e^{-ikx} \right) = \hat{\varphi}(x) + f_2 n_2(x) \qquad (1.17)$$

$$\hat{U}^\dagger \hat{\phi}(x) \hat{U} = \hat{\phi}(x) + f_1 n_1(x); \quad \hat{U}^\dagger \hat{\phi}^\dagger(x) \hat{U} = \hat{\phi}^\dagger(x) + f_1 n_1(x) \qquad (1.18)$$

where,

$$n_1(x) = \frac{2\cos(qx)}{\sqrt{2E_q L}}; \quad n_2(x) = \frac{2\cos(kx)}{\sqrt{2\omega_k L}} \qquad (1.19)$$

This yields,

$$\hat{U}^\dagger :\hat{\phi}^\dagger \hat{\phi}: \hat{\varphi}\hat{U} =: \left( \hat{\phi}^\dagger(x) + f_1 n_1(x) \right)\left( \hat{\phi}(x) + f_1 n_1(x) \right):\left( \hat{\varphi}(x) + f_2 n_2(x) \right) \qquad (1.20)$$

This becomes,

$$\hat{U}^\dagger :\hat{\phi}^\dagger \hat{\phi}: \hat{\varphi}\hat{U} =: \hat{\phi}^\dagger \hat{\phi}: \hat{\varphi} + f_1 n_1 \left( \hat{\phi}^\dagger + \hat{\phi} \right)\hat{\varphi} + f_1^2 n_1^2 \hat{\varphi} + :\hat{\phi}^\dagger \hat{\phi}: f_2 n_2 + f_1 f_2 n_1 n_2 \left( \hat{\phi}^\dagger + \hat{\phi} \right) + f_1^2 f_2 n_1^2 n_2$$
$$(1.21)$$

We also obtain,

$$\hat{U}^\dagger :\hat{\varphi}^4: \hat{U} =: (\hat{\varphi}+f_2 n_2)^4 := \left(:\hat{\varphi}^4: +4f_2 n_2 :\hat{\varphi}^3: +6(f_2 n_2)^2 :\hat{\varphi}^2: +4(f_2 n_2)^3 \hat{\varphi}+(f_2 n_2)^4\right)$$

(1.22)

Note that on the left hand side of the above equations the normal order is done first and then we take the unitary transformation. On the right hand side we have performed the unitary transformation and then applied the normal order. Normally this is not correct, that is, the unitary transformation and normal ordering do not "commute". However in this case it is allowed because the only effect of the unitary transformation is to add a real valued function to the field operator.

Next, consider the state vector defined by,

$$|\Omega'\rangle = \hat{U}|\Omega_v\rangle$$

(1.23)

The energy of this state is given by,

$$\langle\Omega'|\hat{H}|\Omega'\rangle = \langle\Omega_v|U^\dagger \hat{H} U|\Omega_v\rangle$$

(1.24)

Using the above results we will show that $\langle\Omega'|\hat{H}|\Omega'\rangle$ is less than $\langle\Omega_v|\hat{H}|\Omega_v\rangle$ in contradiction of (1.9).

Use (1.1) in (1.24) to obtain,

$$\langle\Omega'|\hat{H}|\Omega'\rangle = \langle\Omega_v|\left(U^\dagger \hat{H}_{0,s} U + U^\dagger \hat{H}_{0,cs} U + \lambda_1 \int U^\dagger :\hat{\varphi}^\dagger \hat{\varphi}: \hat{\varphi} U dx + \lambda_2 \int U^\dagger :\hat{\varphi}^4: U dx\right)|\Omega_v\rangle$$

(1.25)

Note that the integration limits are from $(-L/2)$ to $(L/2)$. Next use (1.15), (1.16), (1.20), and (1.22) in the above to obtain,

$$\langle\Omega_v|\hat{U}^\dagger \hat{H}\hat{U}|\Omega_v\rangle = \langle\Omega_v|\hat{H}|\Omega_v\rangle + f_1 A_1 + f_2(A_2 + B_2) + f_1 f_2 A_3 + f_2^2(\omega_k + B_1)$$
$$+ f_2^3 B_3 + f_2^4 B_4 + f_1^2 (A_4 + f_2 A_5)$$

(1.26)

where,

$$A_1 = E_q \langle\Omega_v|\left[\hat{b}_q^\dagger + \hat{b}_q + \hat{d}_q^\dagger + \hat{d}_q\right]|\Omega_v\rangle + \lambda_1 \int \langle\Omega_v|\left(\hat{\varphi}^\dagger(x)+\hat{\varphi}(x)\right)\hat{\varphi}(x)|\Omega_v\rangle n_1(x) dx \quad (1.27)$$

$$A_2 = \omega_k \langle\Omega_v|\left(\hat{a}_k^\dagger + \hat{a}_k\right)|\Omega_v\rangle + \lambda_1 \int \langle\Omega_v|:\hat{\varphi}^\dagger(x)\hat{\varphi}(x):|\Omega_v\rangle n_2(x) dx \quad (1.28)$$

$$A_3 = \lambda_1 \int \langle\Omega_v|\left(\hat{\varphi}^\dagger(x)+\hat{\varphi}(x)\right)|\Omega_v\rangle n_1(x) n_2(x) dx \quad (1.29)$$

$$A_4 = 2E_q + \lambda_1 \int \langle\Omega_v|\hat{\varphi}(x)|\Omega_v\rangle n_1^2(x) dx \quad (1.30)$$

$$A_5 = \lambda_1 \int n_1^2(x) n_2(x) dx \tag{1.31}$$

$$B_1 = 6\lambda_2 \int n_2^2(x) \langle \Omega_v | \hat{\varphi}^2(x) | \Omega_v \rangle dx \tag{1.32}$$

$$B_2 = 4\lambda_2 \int n_2(x) \langle \Omega_v | \hat{\varphi}^3(x) | \Omega_v \rangle dx \tag{1.33}$$

$$B_3 = 4\lambda_2 \int n_2^3(x) \langle \Omega_v | \hat{\varphi}(x) | \Omega_v \rangle dx \tag{1.34}$$

$$B_4 = 4\lambda_2 \int n_2^4(x) dx \tag{1.35}$$

Select $q$ and $k$ in (1.19) so that $A_5 > 0$. This can be done by setting $k = 2q$ so that $A_5 = \lambda_1 \int (\cos(qx))^2 \cos(2qx) dx > 0$. Next, let $f_2$ be negative so that $f_2 = -|f_2|$. Use this in (1.26) to obtain,

$$\langle \Omega_v | \hat{U}^\dagger \hat{H} \hat{U} | \Omega_v \rangle = \langle \Omega_v | \hat{H} | \Omega_v \rangle + f_1 A_1 - |f_2|(A_2 + B_2) - f_1 |f_2| A_3 + f_2^2 (\omega_k + B_1)$$
$$- |f_2|^3 B_3 + f_2^4 B_4 + f_1^2 (A_4 - |f_2| A_5) \tag{1.36}$$

Make $|f_2|$ large enough so that the term $(A_4 - |f_2| A_5)$ negative. At this point we hold $f_2$ fixed and increase $f_1$. As $f_1 \to \infty$ the last term in (1.36) will be negative and will dominate the other terms. Therefore $\langle \Omega_v | \hat{U}^\dagger \hat{H} \hat{U} | \Omega_v \rangle$ will be less than $\langle \Omega_v | \hat{H} | \Omega_v \rangle$ by an arbitrarily large amount which contradicts (1.9). Therefore there is no lower bound to the energy for the Hamiltonian given by (1.1).

### **3. Conclusion.**

It has been shown that for the toy model Hamiltonian considered in this paper there is no lower bound to the energy. This is consistent with previous work showing the Hamiltonians of the Dirac-Maxwell fields in the temporal or covariant gauges also had no lower bound. The question, then, arises how do we know that a given Hamiltonian has a lower bound or not? This is a particularly important question because the Standard Model of QFT has a number of interacting fields. How do we know that Standard Model has a lower bound to the energy? It is generally assumed that it does but there is no proof, it is a only a conjecture. One purpose of the paper is to show that one should not assume that a given Hamiltonian has a lower bound to the energy. The Hamiltonian in question must be examined to determine that this is the case.